\documentclass[twocolumn,showpacs,amsmath,amssymb,appendix,prd]{revtex4-1}

\usepackage{graphicx}
\usepackage{dcolumn}
\usepackage{bm}
\usepackage{epsfig}
\usepackage{amsmath}
\usepackage{color}
\usepackage{float}

\newcommand{\be}{\begin{equation}}
\newcommand{\ee}{\end{equation}}
\newcommand{\ba}{\begin{eqnarray}}
\newcommand{\ea}{\end{eqnarray}}

\newcommand{\tx}{{\theta_{12}}}
\newcommand{\ty}{{\theta_{13}}}
\newcommand{\tz}{{\theta_{23}}}

\newcommand{\tym}{{\theta_{13}^{m}}}
\newcommand{\dl}{{\Delta m_{31}^2}}
\newcommand{\ds}{{\Delta m_{21}^2}}

\newcommand{\pme}{P_{\mu e}}
\newcommand{\pmm}{P_{\mu \mu}}

\newcommand{\dchsq}{\Delta\chi^2}
\newcommand{\dmm}{\Delta m^2_{\mu\mu}}

\begin{document}
\input{epsf}

\title{Mass hierarchy and CP-phase sensitivity of ORCA using Fermilab Neutrino Beam}

\author{Ushak Rahaman}
\email{ushak@phy.iitb.ac.in}

\affiliation{Department of Physics, 
Indian Institute of Technology Bombay, Powai, Mumbai 400076, India}

\author{Soebur Razzaque}
\email{srazzaque@uj.ac.za}

\affiliation{Department of Physics, University of Johannesburg, PO Box
524, Auckland Park 2006, South Africa}

\begin{abstract} 
We explore neutrino mass hierarchy determination and CP-phase measurement 
using multi-megaton water Cherenkov detector KM3NeT-Oscillation Research
with Cosmics in the Abyss (ORCA) in the Mediterranean sea receiving neutrino 
beam from the Fermilab Long Baseline Neutrino Facility (LBNF) over 6900 km baseline.  
We find that with the proposed beam luminosity of $1.2\times 10^{21}$ proton-on-target per year, it will be possible to determine mass hierarchy at $\gtrsim 4\,\sigma$ confidence level within 1 year in the neutrino mode alone. A combined 1 year in neutrino and 1 year in antineutrino mode can determine hierarchy at $\gtrsim 6\,\sigma$ confidence level.  
We also find that a non-zero CP-phase can be detected with up to 
$\sim 1.8\,\sigma$ significance after 10 years of data taking.  
We explore degeneracy of neutrino oscillation parameters and uncertainties 
in detection efficiencies affecting the results.     
\end{abstract}

\pacs{14.60.Lm, 14.60.Pq}

\date{\today}
\maketitle

\section{Introduction}

Determination of neutrino mass hierarchy and measurement of the 
CP-phase have been widely discussed in recent years using multi-megaton
water/ice Cherenkov detectors and with atmospheric neutrino 
flux~\cite{Mena:2008rh, Akhmedov:2012ah, Winter:2013ema, Ribordy:2013set, 
Franco:2013in, Aartsen:2014oha, Razzaque:2014vba, Adrian-Martinez:2016fdl,Ge:2013zua, 
Ge:2013ffa, Capozzi:2015bxa, Adrian-Martinez:2016zzs}. 
Such measurements are based on neutrino propagation over long-baseline, 
through the mantle and core of the earth, for which oscillations in vacuum 
and in matter including MSW resonance~\cite{msw1, Mikheev:1986gs} and parametric 
enhancement effects~\cite{Akhmedov:1998ui, Liu:1998nb} are relevant. Determining 
the mass hierarchy and measuring the CP-phase are among the foremost objectives of 
neutrino physics as well as particle physics. 

While atmospheric neutrinos can provide large statistics of events in multi-megaton 
scale water/ice Cherenkov detectors with few GeV threshold, such as the proposed 
``Precision IceCube Next Generation Upgrade (PINGU)''~\cite{Aartsen:2014oha} and 
``Oscillation Research with Cosmics in the Abyss (ORCA)" option of 
KM3NeT~\cite{Adrian-Martinez:2016fdl}, there are a number of uncertainties 
which reduce the significance to hierarchy determination and CP-phase measurement.  
The uncertainties directly related to atmospheric neutrinos are flux normalizations 
and spectral shapes, presence of both neutrinos and antineutrinos in the flux, 
presence of both $\nu_e$ and $\nu_\mu$ flavors in the flux, etc.  Careful analyses 
have been carried out to minimize these uncertainties with dedicated 
detector-specific simulations, resulting $\approx 3\sigma$ significance 
for hierarchy determination within $\approx 3-4$ years of data 
taking~\cite{Kouchner:2016pqa, TheIceCube-Gen2:2016cap}.  Sensitivity to the
CP-phase $\delta$ is rather poor unless further upgrades in future is 
considered~\cite{Razzaque:2014vba}, giving $\sim 1$ megaton effective mass at sub-GeV energies. 

Long baseline accelerator experiments such as T2K and NO$\nu$A currently 
provide no significant constraint on hierarchy, with only a weak preference 
to normal mass hierarchy~\cite{Abe:2015awa, Adamson:2016xxw}.  
Sensitivities to CP-phase are similarly poor, with $90\%$ C.L.\ exclusion of 
the $\delta = (0.15, 0.83)\pi$ region for normal hierarchy and 
$\delta = (-0.08, 1.09)\pi$ region for inverted hierarchy by 
T2K ~\cite{Abe:2015awa}.  NO$\nu$A excluded $\delta = (0.1, 0.5)\pi$ 
region for inverted hierarchy at $90\%$ C.L.\ most recently by using 
$\nu_e$ appearance data~\cite{Adamson:2016tbq}.  Future accelerator 
experiment such as ``Deep Underground Neutrino Experiment (DUNE)'' 
using neutrino beam provided by ``Long Baseline Neutrino Facility (LBNF)'' 
at the Fermilab is expected to determine hierarchy and possibly find 
evidence of a non-zero CP-phase~\cite{Acciarri:2016crz}. 

A combination of multi-megaton scale water/ice Cherenkov detector at the 
South Pole receiving accelerator neutrino beam from the Northern hemisphere 
has been explored recently to do oscillation physics~\cite{Fargion:2010vb, Tang:2011wn}.  
In particular sensitivities to mass hierarchy and CP-phase was explored by choosing 
a specific example from the CERN to PINGU over a baseline of 11810 km~\cite{Tang:2011wn}. 
Possibilities of exploring mass hierarchy in experimental set ups with beam from 
CERN (Fermilab) to Lake Baikal (KM3NeT)~\cite{Agarwalla:2012zu} and from CERN to Super-Kamiokande~\cite{LujanPeschard:2013ud} have been explored as well. Other, shorter, baselines have also been explored using neutrino beam and huge atmospheric detectors to determine mass hierarchy~\cite{Brunner:2013lua, Vallee:2016xde}.
 
In this paper we explore neutrino mass hierarchy and CP-phase sensitivities of 
the KM3NeT-ORCA detector in the Mediterranean sea, receiving LBNF neutrino beam 
over a baseline of 6900 km.  We use proposed characteristics of the detector from 
simulations~\cite{Adrian-Martinez:2016fdl} such as energy resolution and efficiency 
to specific interaction channels.  Similarly we use neutrino and antineutrino beam 
characteristics of the LBNF beam~\cite{Acciarri:2016crz}.  We investigate hierarchy 
sensitivity with marginalizations over all relevant oscillation parameters.  We also investigate degeneracy of the oscillation parameters to measure the CP-phase.  We address particle flavor (electrons and muons) misidentification in the detector affecting our results.   

We discuss briefly long baseline oscillation relevant for Fermilab to 
ORCA in Sec.~\ref{oscillation}.   In Sec.~\ref{beamdetector} we discuss
LBNF beam and detector characteristics as well as details of simulations.  
We present and discuss our results in Sec.~\ref{results} and draw conclusions 
in Sec.~\ref{conclude}.
 
\section{Long baseline oscillation}
\label{oscillation}

The probability $\pme$ of $\nu_{\mu}$ oscillating into $\nu_e$ and the $\nu_\mu$ survival probability 
$P_{\mu\mu}$ are functions of mixing angles $\tx$, $\ty$, $\tz$, mass-squared differences $\dl$ and 
$\ds$ for 3 flavors and the CP-violating phase $\delta$. The baseline for the LBNF beam from the 
Fermilab to ORCA site, off the coast of Toulon, France~\cite{Adrian-Martinez:2016fdl} is $L=6900$~km. 
Neutrinos along this trajectory do not pass through the core of the earth and a quasi constant density 
approximation is reasonable for calculation of oscillation probabilities.  Sensitivity to the mass 
hierarchy comes mostly at energies $\gtrsim 2$~GeV and one can ignore 1-2 mixing and mass splitting 
in this energy range.  The corresponding neutrino oscillation probabilities for normal mass hierarchy 
(NH) are~\cite{Akhmedov:1998xq, Akhmedov:2008qt, Akhmedov:2012ah}
\ba
P_{\mu e} &=& \sin^2\theta_{23} P_A \nonumber \\
P_{\mu\mu} &=& 1- \frac{1}{2} \sin^2 2\theta_{23} - \sin^4 \theta_{23} P_A \nonumber \\
&& + \frac{1}{2} \sin^2 2\theta_{23} \sqrt{1-P_A} \cos\phi_X ~,
\label{oscHigh}
\ea
where, for constant density approximation,
\ba
P_A &=& \sin^2 2\theta_{13}^m \sin^2 \frac{\phi_{31}^m}{2} \nonumber \\
\phi_X &=& \tan^{-1} \left( \cos 2\theta_{13}^m \tan \frac{\phi_{31}^m}{2} \right) \nonumber \\
&& +  \frac{\dl L}{4E_\nu}  \left( 1 + \frac{2VE_\nu}{\dl} \right)
\label{PAphiX}
\ea 
and
\ba
\phi_{31}^m &=& \frac{\dl L}{2E_\nu} {\sqrt{ \left( \cos2\ty - \frac{2VE_\nu}{\dl} \right)^2 + \sin^22\ty}}  \nonumber \\
\cos2\tym &=& \frac{\cos2\ty - \frac{2VE_\nu}{\dl}} {\sqrt{ \left( \cos2\ty - \frac{2VE_\nu}{\dl} \right)^2 + \sin^22\ty}}
\label{matter31}
\ea
are the oscillation phase and mixing angle in matter, respectively.  The matter potential $V = \sqrt{2} G_F N_e$, with $G_F$ and $N_e$ being the Fermi constant and average electron number density along the trajectory, respectively.

Oscillation probabilities for antineutrinos ${\bar P}_{\mu e}$ and ${\bar P}_{\mu\mu}$ can be written by substituting $V \to -V$ in the above equations.  It can be seen from Eqs.~(\ref{matter31}) and (\ref{PAphiX}) that for antineutrinos  $\cos2\tym \approx 1$ and correspondingly $P_A (V \to -V)  \approx 0$.  Therefore the antineutrino probability ${\bar P}_{\mu e}$ in matter is suppressed.  Also, for antineutrinos $\phi_X (V\to -V) \approx \phi_{31} = \dl L/2E_\nu$ and the probability ${\bar P}_{\mu\mu}$ assumes the vacuum oscillation probability. 

For inverted mass hierarchy (IH) the oscillation probabilities for the neutrino and antineutrinos are exchanged in this approximation of zero 1-2 mixing such that $P^{\rm IH}_{\alpha\beta} = {\bar P}^{\rm NH}_{\alpha\beta}$ and ${\bar P}^{\rm IH}_{\alpha\beta} = P^{\rm NH}_{\alpha\beta}$~\cite{Akhmedov:2012ah}.

\begin{figure}[t]
\centering
\includegraphics[width=0.45\textwidth]{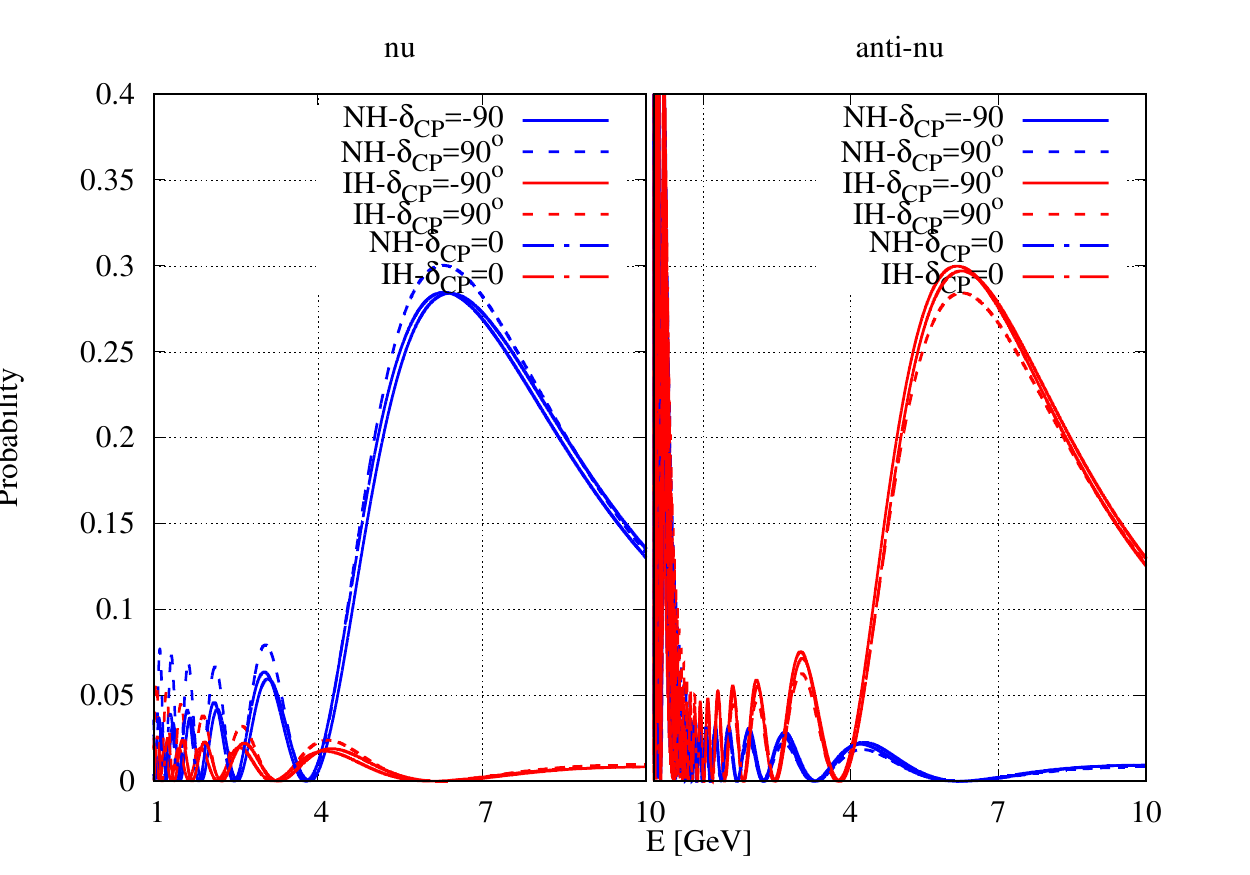}
\includegraphics[width=0.45\textwidth]{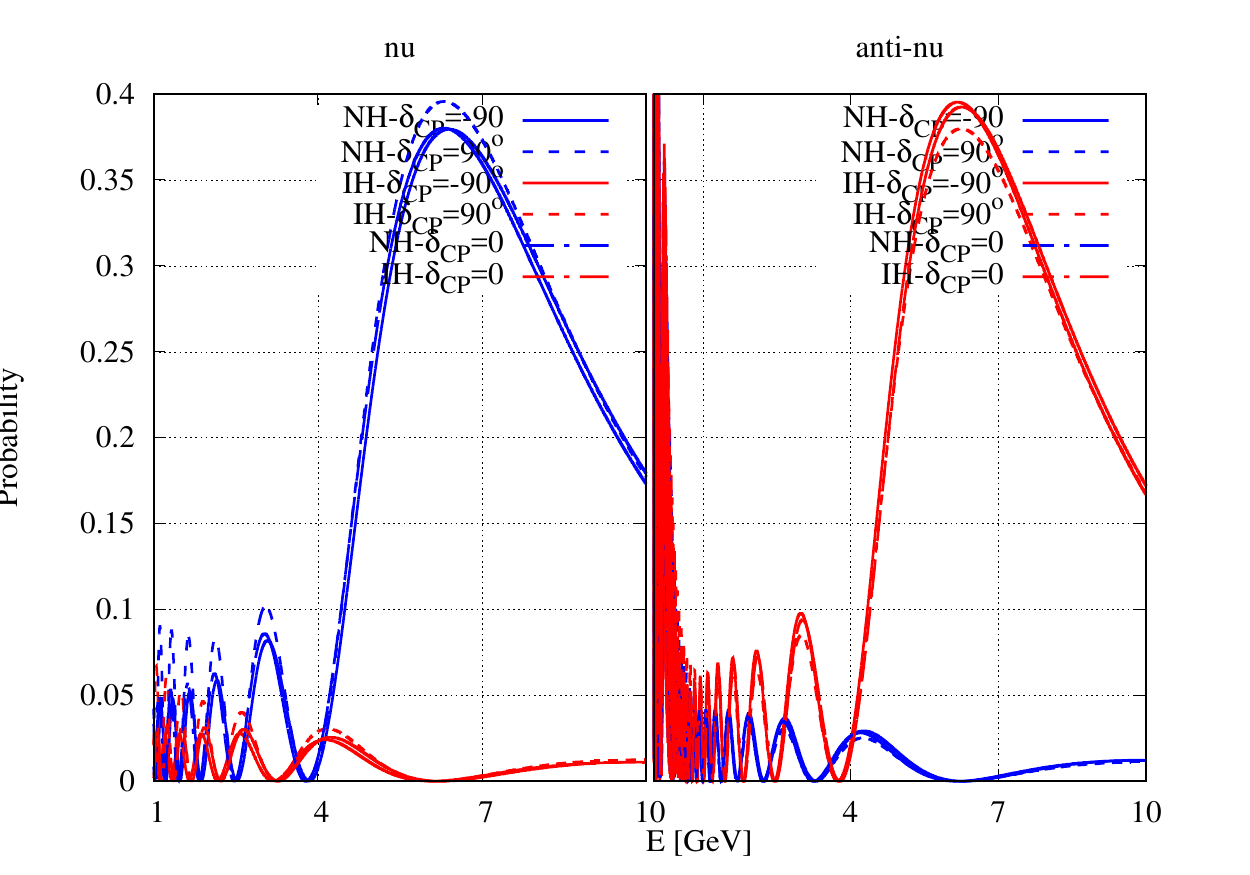}
\caption{\footnotesize{Probabilities $\pme$ (${\bar P}_{\mu e}$) in the left (right) panels as functions of energy for different hierarchies and $\delta$ values for a baseline of 6900 km. The upper (lower) panels show the probabilities for $\sin^2 \tz=0.44$ (0.584).}}
\label{prob1me}
\end{figure}

Our numerical calculation of  probabilities, using the Preliminary Reference Earth Model~\cite{Dziewonski:1981xy}, 
agree very well with above expectation for NH and IH for neutrinos and antineutrinos, for the same CP-phase values.  
We have used the best fit values of the neutrino oscillation parameters from Refs.~\cite{Esteban:2016qun, nufit}
for numerical calculation.

In Fig.~\ref{prob1me}, we have plotted $\pme$ (left panels) and ${\bar P}_{\mu e}$ (right panels) 
for NH and IH and for three values of $\delta$.  The upper and lower panels correspond to the lower 
octant (LO) and higher octant (HO) of $\tz$, respectively.  We have used a baseline of 6900 km 
(distance between Fermilab and the ORCA detector in the Mediterranean sea).  The dominant feature
of the probabilities in these plots is the 1-3 resonance at an energy 
$E^{\rm R}_\nu = \dl \cos 2\ty/2V \approx 6.6$~GeV for NH (IH) in case of $\pme$ (${\bar P}_{\mu e}$).  
The sensitivity to hierarchy mostly comes from this feature in this energy region. 
The effect of octant for $\tz$ is just a shift in the overall probability values since those 
are proportional to $\sin^2 \tz$ as in Eq.~(\ref{oscHigh}).  It is clear from Fig.~\ref{prob1me} 
that unlike other long baseline experiments such as NO$\nu$A and T2K, no hierarchy-$\delta$ degeneracy
\cite{Mena:2004sa,Prakash:2012az} exists for this baseline and the probabilities for NH is well separated from IH. 

\begin{figure}[t]
\centering
\includegraphics[width=0.45\textwidth]{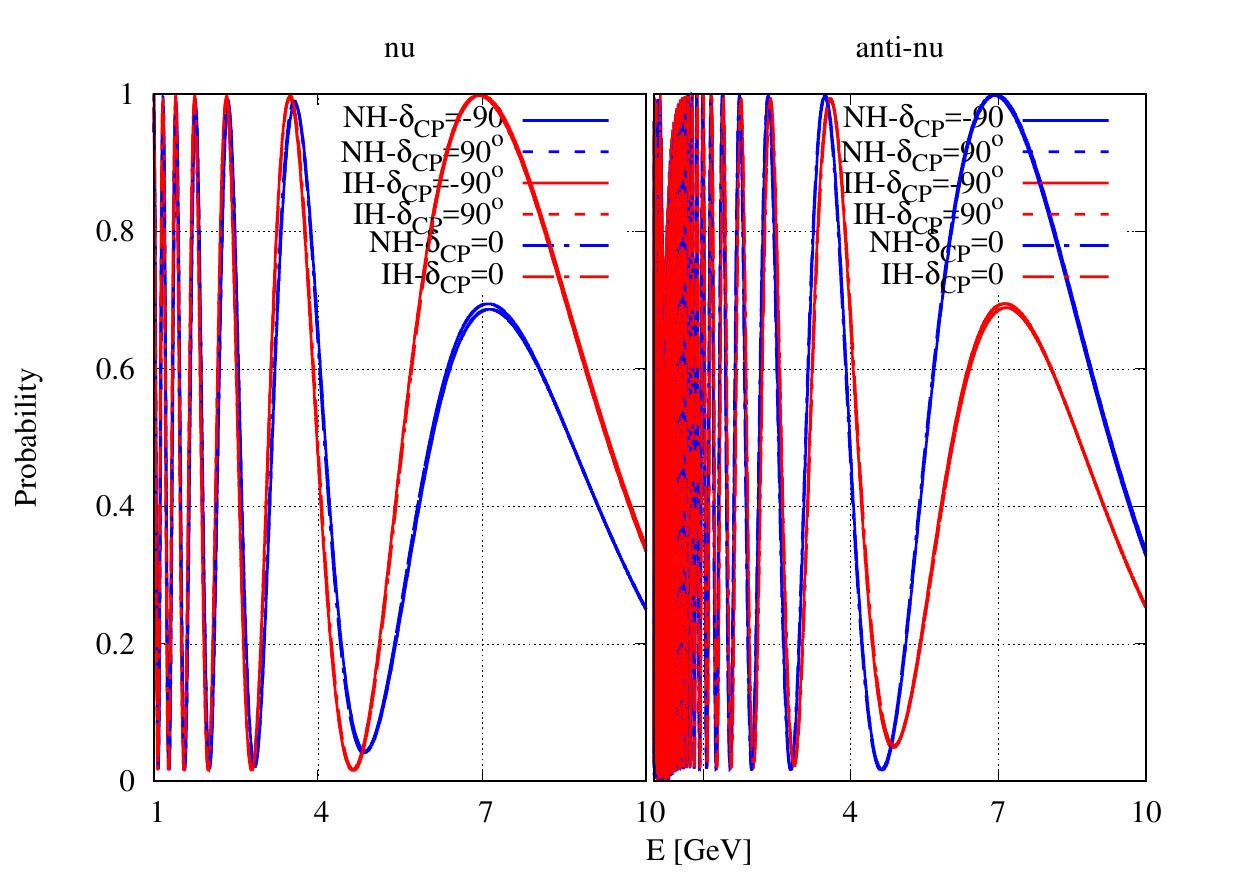}
\includegraphics[width=0.45\textwidth]{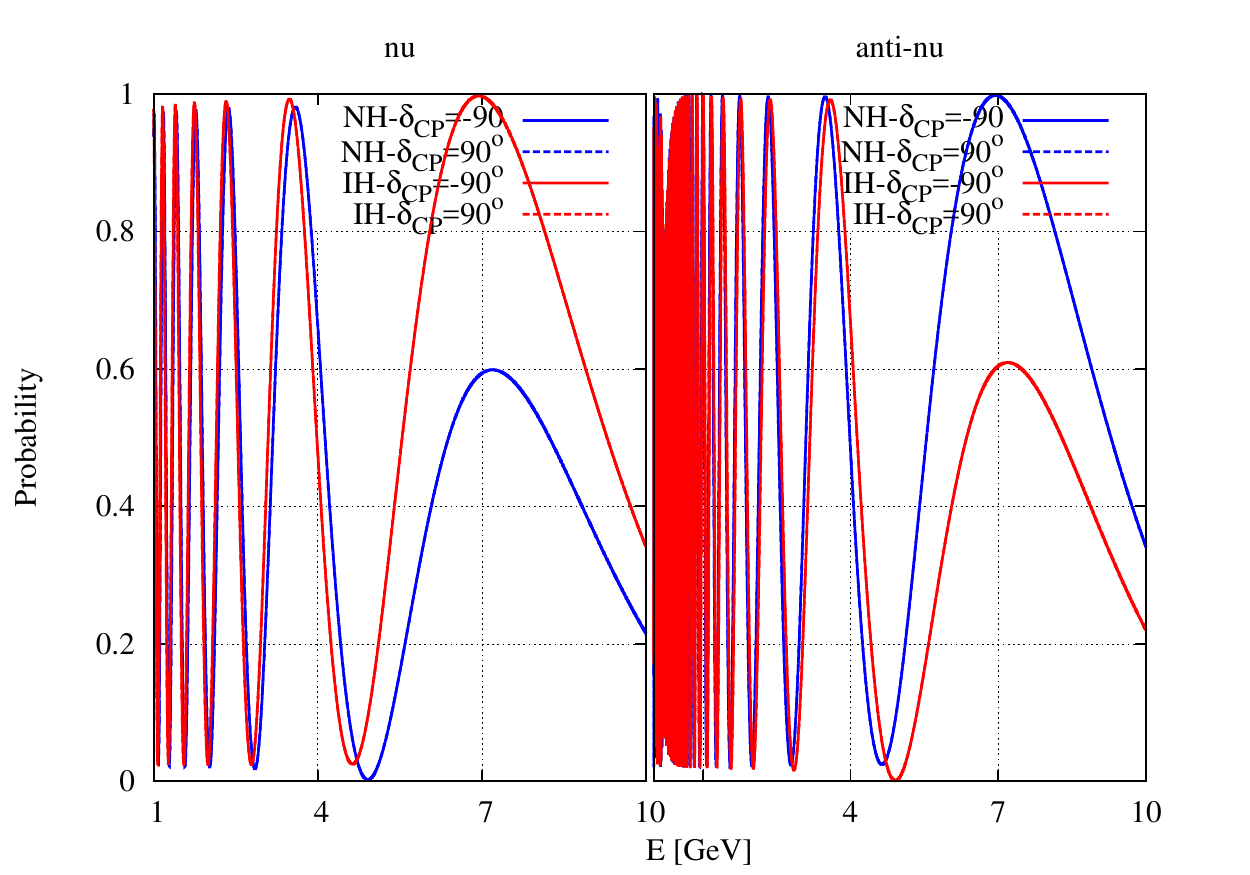}
\caption{\footnotesize{The same as Fig.~\ref{prob1me} but for $\pmm$ and ${\bar P}_{\mu\mu}$.}}
\label{prob2mm}
\end{figure}

We have plotted the survival probabilities $\pmm$ (left panels) and ${\bar P}_{\mu\mu}$ (right panels) in Fig.~\ref{prob2mm}.  As discussed for Eqs.~(\ref{oscHigh})-(\ref{matter31}), the probabilities ${\bar P}^{\rm NH}_{\mu\mu} = P^{\rm IH}_{\mu\mu}$ are consistent with vacuum oscillation, which are different from the probabilities  ${\bar P}^{\rm IH}_{\mu\mu} = P^{\rm NH}_{\mu\mu}$.  These differences give sensitivity to hierarchy in the muon neutrino channel.  Again, the upper and lower panels in Fig.~\ref{prob2mm} correspond to LO and HO of $\tz$ as in Fig.~\ref{prob1me}.  But the difference between the LO and HO for $\pmm$ is opposite than for $\pme$, as evidenced from Eq.~(\ref{oscHigh}).

To understand the CP-dependence of the probabilities one cannot ignore the 1-2 mixing however, as was done in Eq.~(\ref{oscHigh}).  The 1-2 oscillation phase at low energies and close to the 1-2 resonance energy in constant density medium is~\cite{Razzaque:2014vba}
\ba
\phi_{21}^m &=& \frac{\ds L}{2E_\nu} {\sqrt{ \left( \cos2\tx - \frac{2VE_\nu}{\ds} \right)^2 + \sin^22\tx}} .
\label{phi21}
\ea
The CP-dependent parts of the probabilities, after averaging over fast oscillations driven by the 1-3 oscillation phase are~\cite{Razzaque:2014vba}
\ba
\langle P_{e\mu}^\delta \rangle &=& \frac{J_\theta}{2} 
\left[
\cos\delta \cos 2\theta_{12}^m \sin^2 \frac{\phi_{21}^m}{2} + 
\frac{1}{2} \sin\delta \sin \phi_{21}^m
\right]
\nonumber \\
\langle P_{\mu\mu}^\delta \rangle &=& - \frac{J_\theta}{2} \cos\delta \sin^2 \frac{\phi_{21}^m}{2} 
\cos 2\theta_{12}^m
\label{oscCP}
\ea
where
\ba
J_\theta \equiv \sin 2\tz \sin 2\theta_{12}^m \sin 2\tym \cos\tym
\ea
is the Jarlskog invariant in matter.  Note that $P_{\mu e}^\delta = P_{e\mu}^{-\delta}$ and the probabilities for antineutrinos can be found by changing $V \to -V$ and $\delta \to -\delta$ in Eq.~(\ref{oscCP}). For energies above the 1-2 resonance, $\cos 2\theta_{12}^m \approx -1$.

Equation~(\ref{oscCP}) describes the general behavior of the CP-dependence of the numerically calculated probabilities in Figs.~\ref{prob1me} and \ref{prob2mm} rather well.  A systematic shift in $\pme$ with $\delta$ over a large energy range in Fig.~\ref{prob1me}  is due to the CP-odd second term of the first line in Eq.~(\ref{oscCP}) containing $\sin\delta$, as discussed in details in Ref.~\cite{Razzaque:2014vba}.  The main contribution to the CP sensitivity comes from this term in $\pme$.  The effect of CP in case of $\pmm$ is opposite to that of $\pme$ and is CP-even because of the $\cos\delta$ term.  Therefore in Fig.~\ref{prob2mm} the $\delta = 90^\circ$ and $-90^\circ$ probabilities are the same.

\begin{figure}[t]
\centering
\includegraphics[width=0.45\textwidth]{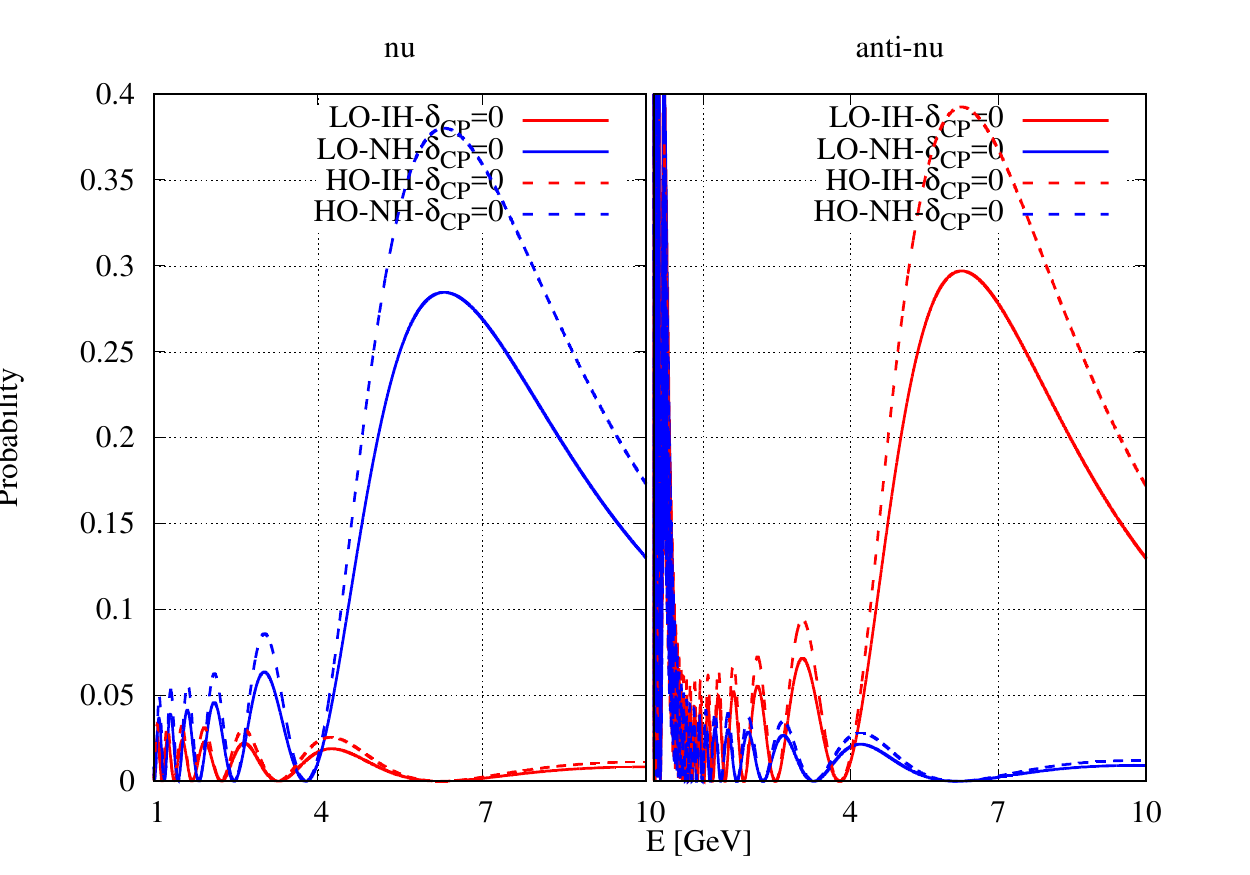}
\includegraphics[width=0.45\textwidth]{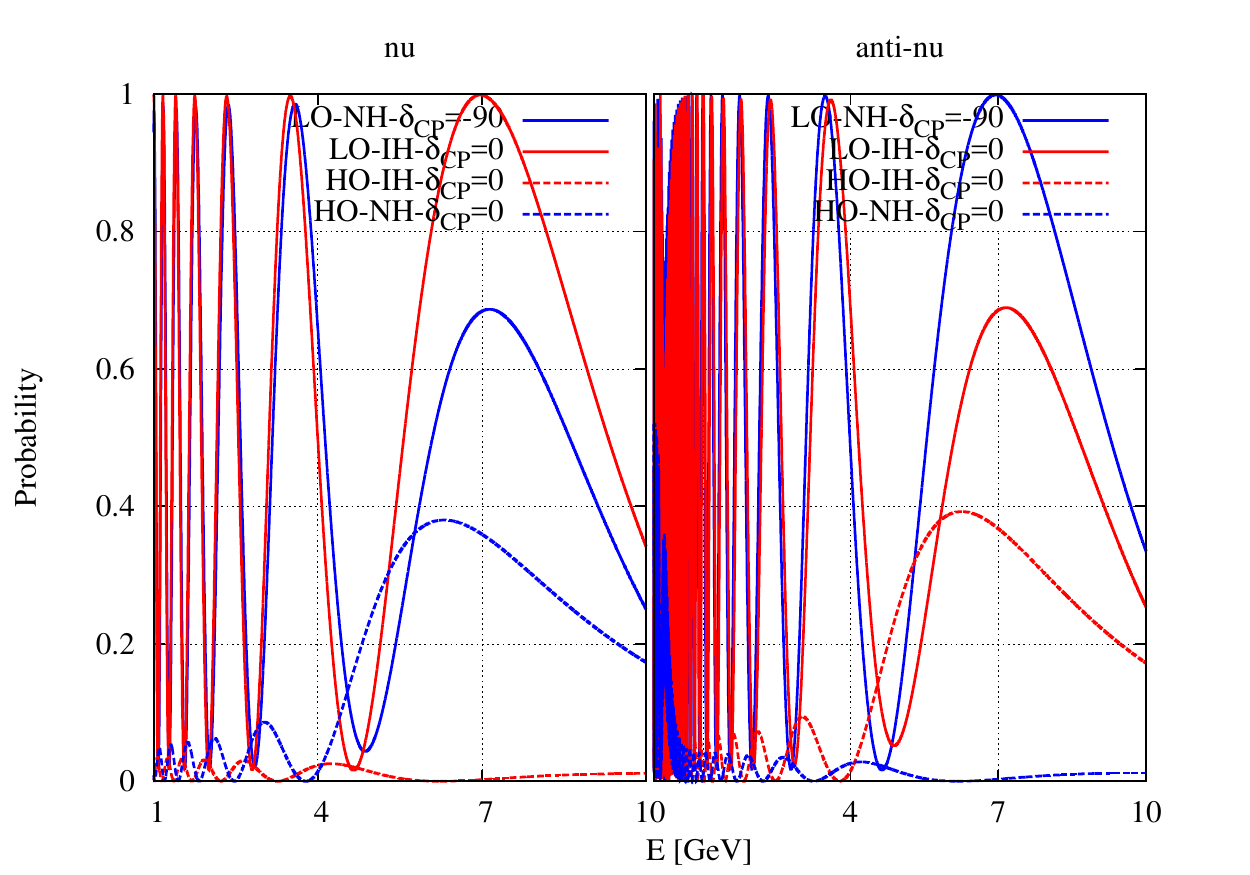}
\caption{\footnotesize{Probabilities $\pme$ ($\pmm$) in the upper (lower) panels for different octants of $\tz$ and hierarchy combinations but with the same CP phase.  The left (right) panels show the probabilities for neutrino (anti-neutrino).  We have used a baseline of 6900~km.}}
\label{prob7}
\end{figure}

From the above probability plots, we can conclude that since the probabilities for NH and IH are quite well separated, it is possible to get a good hierarchy sensitivity for the proposed Fermilab beam to the ORCA detector in the Mediterranean sea.  The effect of CP on the other hand is rather mild at the probability level and will be difficult to measure.

Next we investigate the effect of marginalization over $\tz$ on the hierarchy sensitivity from the probability plots.  Figure~\ref{prob7} shows the probabilities for different $\tz$ values in the LO and HO, i.e., $\sin^2\tz = 0.440$ and 0.584, respectively.  Both the $\pme$ and $\pmm$ probabilities for different octants and hierarchy combinations are relatively well-separated from each other.  Thus the octant degeneracy cannot mimic a NH probability for IH and vice-versa, and the sensitivity to hierarchy is not lost.

\begin{figure}[t]
\centering
\includegraphics[width=0.45\textwidth]{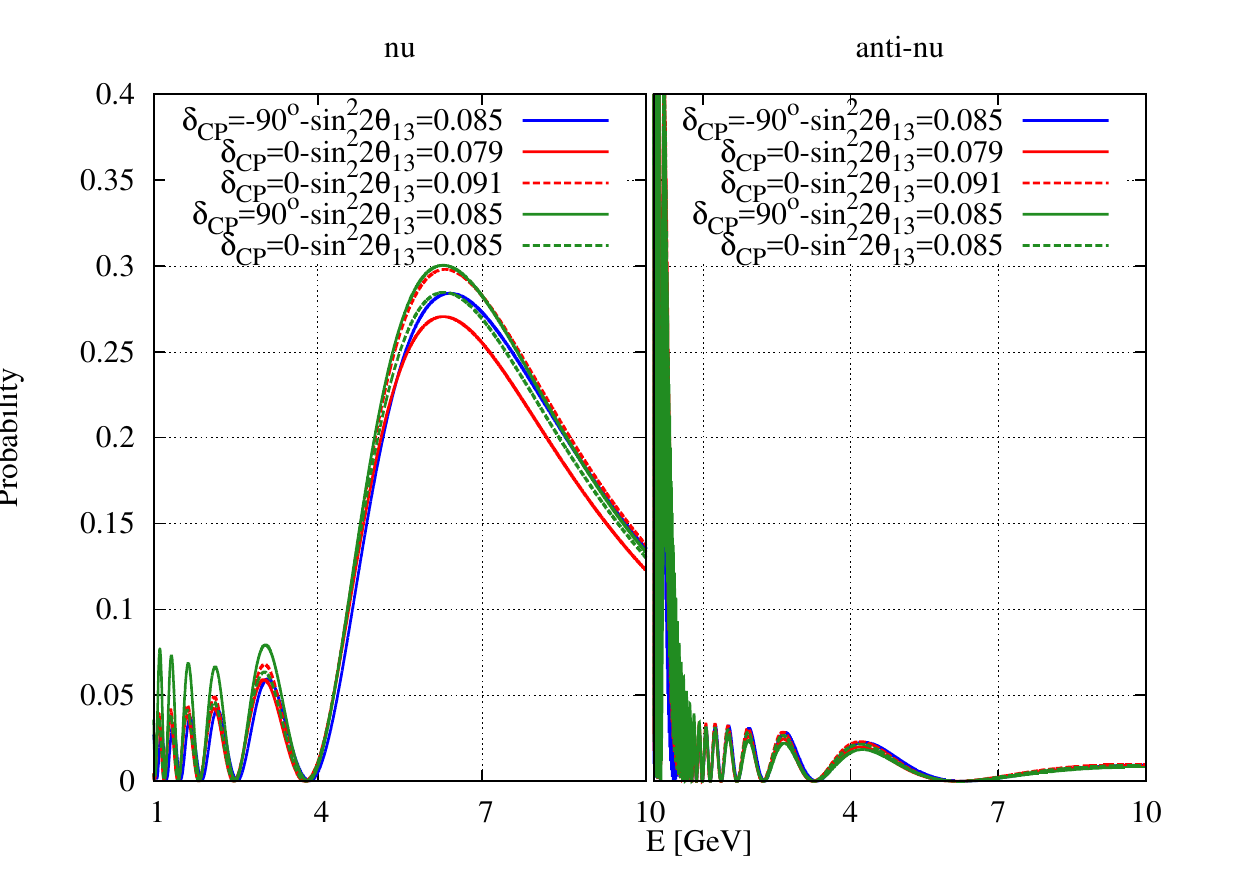}
\includegraphics[width=0.45\textwidth]{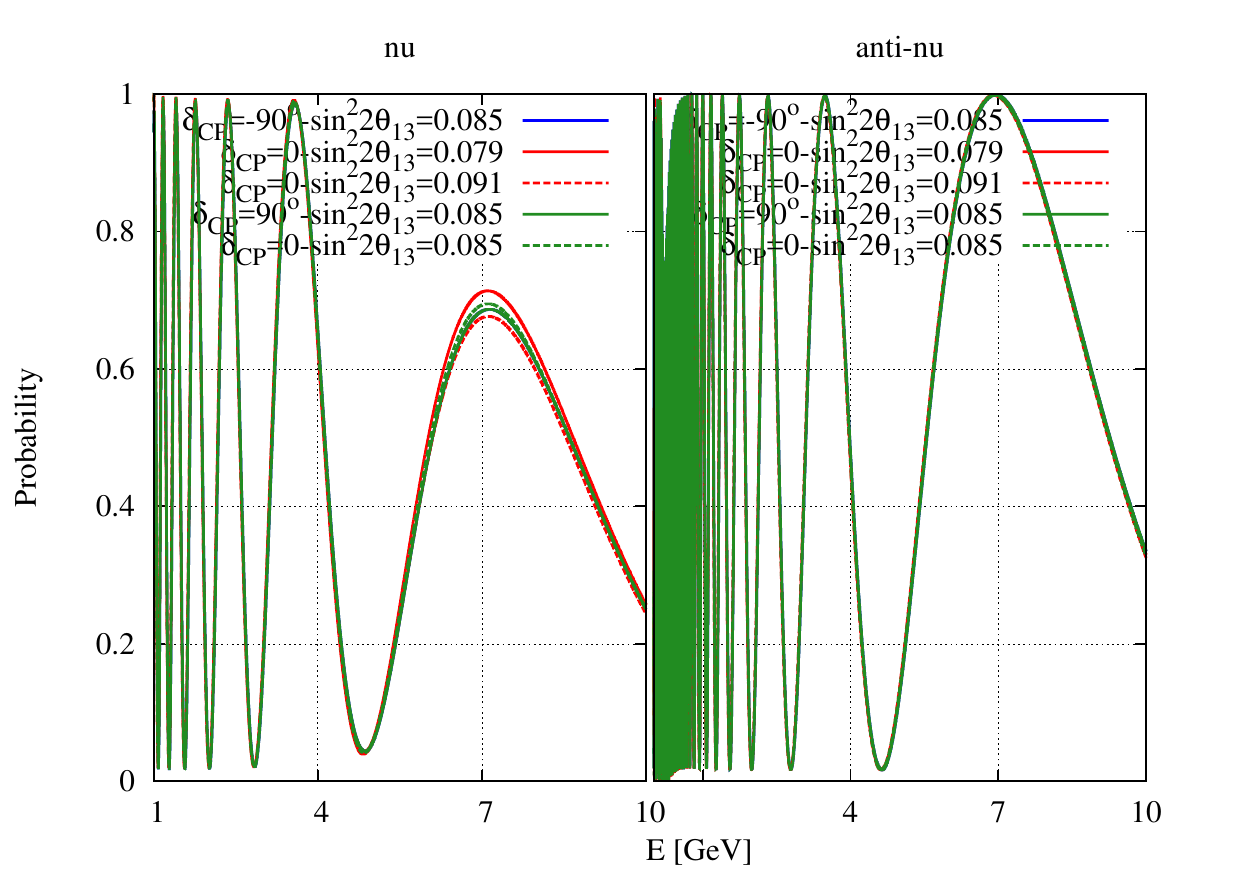}
\caption{\footnotesize{Probabilities $\pme$ ($\pmm$) in upper (lower) panels for LO-NH and different $\delta$-$\sin^22\ty$ combinations. The left (right) panels show the probabilities for neutrino (anti-neutrino).  We have used a baseline of 6900~km.}}
\label{prob8}
\end{figure}

Finally, in Fig.~\ref{prob8} we have plotted the probability curves for LO, NH and different $\delta$ and $\sin^2 2\ty$ combinations for both oscillation and survival cases to investigate the effect of marginalization over $\sin^22\ty$ on the CP-phase sensitivity. In case of $\pme$, the probability for $\sin^2 2\ty=0.085$ and $\delta=-90^\circ$ is the same with the probability for $\sin^2 2\ty=0.085$ and $\delta=0$ and close to all other $\sin^2 2\ty$ and $\delta$ combinations. Again, the probabilities for $\sin^2 2\ty=0.085$ and $\delta=90^\circ$, is very similar to that for $\sin^2 2\ty=0.091$ and $\delta=0$ and close to probabilities with other $\delta$ and $\sin^2 2\ty=0.091$ combinations. In case of survival probabilities, curves with all $\sin^2 2\ty$ and $\delta$ combinations are very close to each other.  Therefore marginalization over $\sin^22\ty$ is expected to reduce sensitivity to the CP-phase.

\section{Beam and detector characteristics and simulation details}
\label{beamdetector}

To do our simulations, we have used the proposed LBNF beam, designed for the
DUNE program~\cite{Acciarri:2016crz}. The beam is bended $7.2^\circ$ westward and 
$5.8^\circ$ downward along the far detector at the Sanford Underground Research 
Facility (SURF) 1300~km away~\cite{Strait:2016mof}. For the location of the
KM3NeT-ORCA detector, the beam will need to be bent $32.8^\circ$ downward. For the ORCA detector, the beam line needs to be turned. In this case DUNE and ORCA cannot share the same beam, but can use the same flux and normalization. We have used 120 GeV beam of 1.2 MW power, corresponding to $1.2\times10^{21}$ protons on target per year.  The dimensions of the muon decay path is 250~m $\times$ 4~m and the beam peaks at around 2.5~GeV. The $\nu_\mu$ beam is contaminated by $\nu_e$, ${\bar \nu}_\mu$ and ${\bar \nu}_e$.

We also use 7 energy bins in the range 1-7 GeV to calculate event rates.
The bin widths are 0.27 GeV, 0.73 GeV, 0.75 GeV, 1 GeV, 1.25 GeV, 1.3 GeV and 0.7 GeV, respectively. Typical number of signal electron (muon) neutrino events in the 1-7 GeV energy range is $\approx 140$/yr ($\approx 630$/yr).
For detector characteristics, we followed specification published in the KM3NeT Letter of 
Intention~\cite{Adrian-Martinez:2016fdl}. The mass of the detector is 3.6~Mton.  
The signal and background efficiencies for different interactions were calculated 
by taking the ratio of the effective volume, given in figure~60 of Ref.~\cite{Adrian-Martinez:2016fdl} 
with the total detector volume of $3.6\times10^6\ \rm m^3$ in the relevant energy range.  Note that this effective volume is at the trigger level and may change with different quality cuts, although the ratios may not change. With this caveat in mind we have calculated efficiencies from the above mentioned figure for different interactions inside the detector for each of the 7 energy bins in the 1-7 GeV range.  These interactions are $e^-$, $e^+$, $\mu^-$ and $\mu^+$ events as well as $\nu_\mu$ and $\bar{\nu}_\mu$ neutral current events.

Signals for the oscillation channel come from the electrons generated by the 
$\nu_\mu \rightarrow \nu_e$ oscillation and backgrounds come from muon misidentification, 
beam contamination and neutral current pion interactions.  Similarly, for the survival channel, 
signals come from muons due to the $\nu_\mu \rightarrow \nu_\mu$ survival and backgrounds come 
from beam contamination and electron misidentification.  We assumed a nominal value of $30\%$ 
for the misidentification factor \cite{Adrian-Martinez:2016fdl} and then changed it to $10\%$ 
to see the effect on the CP phase sensitivity, and to $10\%$ and $50\%$ to see the effect on mass hierarchy sensitivity.

To calculate simulated experimental event rates and theoretical event rates, 
and then to calculate $\dchsq$ from those, we have used the software 
GLoBES~\cite{Huber:2004ka, Huber:2007ji}.  We used the best fit values of 
the neutrino oscillation parameters from Refs.~\cite{Esteban:2016qun, nufit}.  
For LO, the best fit value of $\sin^2\tz$ is 0.44 and for HO, it is 0.584.  
The best fit value of $\sin^2 2\ty$ is 0.085 and that of $\sin^2 \tx$ is 0.31. 
For the solar mass-squared difference we have used $\ds=7.49\times10^{-5}\ \rm eV^{2}$, 
and for the atmospheric mass-square difference, we have used $\dmm=2.52\times10^{-3}\ \rm eV^2 $, 
where~\cite{Nunokawa:2005nx}
\ba
\dmm &=& \sin^2\tx \dl + \cos^2 \tx \Delta m^{2}_{32} \nonumber \\
&& + \cos \delta \sin 2\tx \sin \ty \tan \tx \ds.
\ea

To generate theoretical event rates, we vary the test values of $\sin^2 2\ty$ 
and $\dmm$ in their $2\,\sigma$ range with $3.5\%$ and $3\%$ uncertainty, 
respectively.  We also vary the test values of $\sin^2\tz$ in its $3\,\sigma$ 
range $[0.35, 0.65]$ in steps of 0.01. For the test values of $\delta$, we vary
it in the full range of $[-180^\circ, 180^\circ]$ in steps of $2.5^\circ$. 
We used Gaussian priors for $\sin^2 2\tz$, $\sin^2 2\ty$ and $|\dmm |$.

%




Automatic bin-based energy smearing is done for generated events with a Gaussian smearing function in GLoBES, defined as
\be
R^c(E,E^\prime)=\frac{1}{\sigma(E)\sqrt{2\pi}}e^{-\frac{(E-E^\prime)^2}{2\sigma^2(E)}},
\ee
where $E^\prime$ is the reconstructed neutrino energy.  The corresponding energy resolution function is given by 
\be
\sigma(E)= \alpha E + \beta \sqrt{E} + \gamma ~,
\ee
where we have used $\alpha=\beta=0$ and $\gamma=0.4$ for 
the 1-7 GeV energy range relevant for the Fermilab-LBNF beam.   

For systematics, We have used signal normalization
and signal calibration errors in energy as $5\%$ and $0.01\%$, respectively, for
the $\nu_e$ appearance events.  For the same $\nu_e$ appearance
events the background normalization and calibration errors
are $10\%$ and $0.01\%$, respectively. For the $\nu_\mu$ events
the signal normalization and calibration errors in energy are $2.5\%$
and $0.01\%$ while the background normalization and calibration
errors are $10\%$ and $0.01\%$, respectively. 
Use of systematics in GLoBES has been discussed in details in Refs.~\cite{Huber:2004ka, Huber:2007ji}.

\section{Results and Discussion}
\label{results}

The results from our study are shown in the mass hierarchy and CP-phase 
sensitivity plots in this Section.  We have taken into account various 
uncertainties in the oscillation parameters as well as uncertainties in
the event reconstruction at the detector.  The square-root of the $\Delta\chi^2$
values reported here is the significance in number of $\sigma$ or confidence level (C.L.) for 1-parameter.

\subsection{Mass Hierarchy Sensitivity}

The KM3NeT-ORCA has very good sensitivity to neutrino mass hierarchy, thanks to 
its large mass at an energy range where matter effect is significant for the 
baseline we have considered.  As mentioned earlier, the effect of CP-phase
uncertainty on determination of mass hierarchy is rather mild in this energy range.  

\begin{figure}[thb]
\centering
\includegraphics[width=0.48\textwidth]{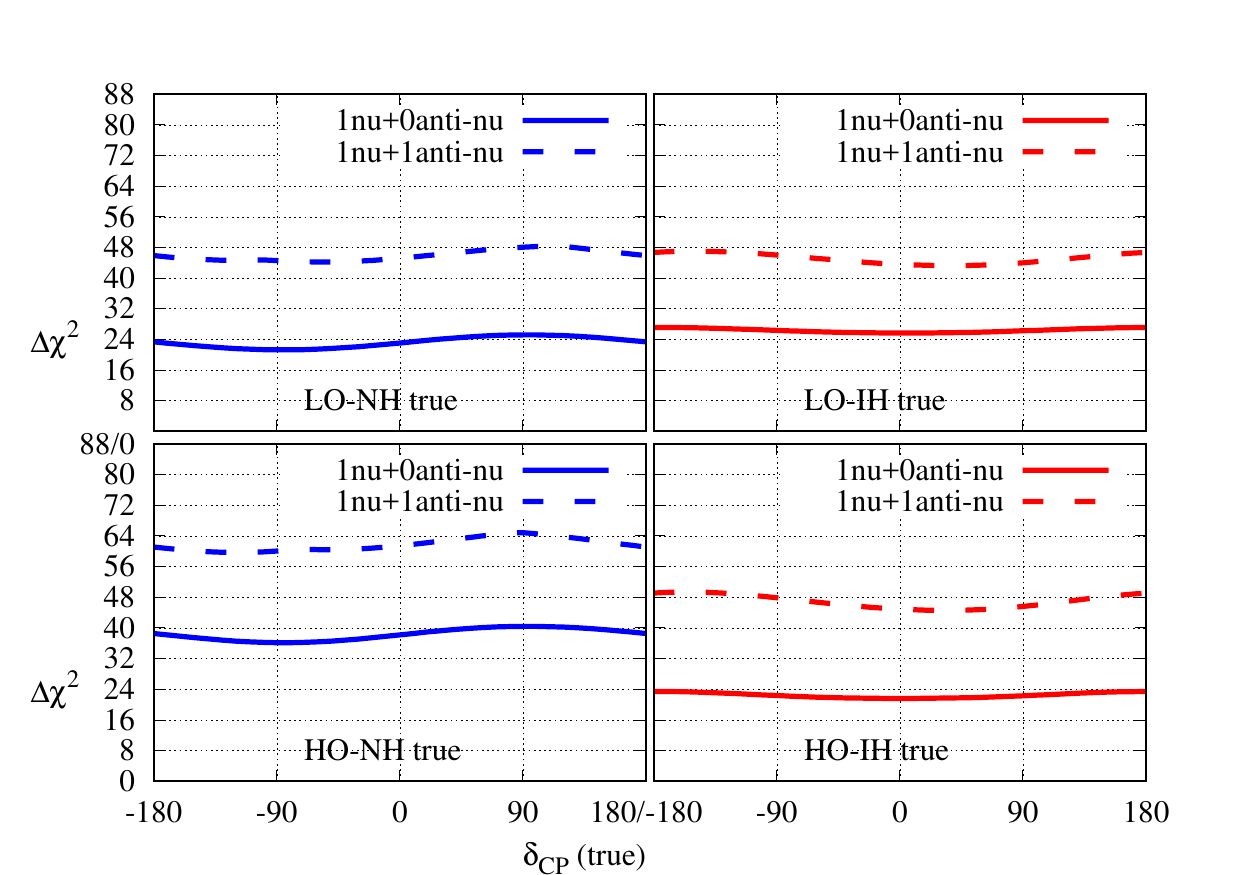}
\caption{\footnotesize{Mass hierarchy discovery potential after 1-year $\nu$ run only (solid lines) and after 1-year $\nu$ and 1-year $\bar{\nu}$ run (dashed lines).  The upper (lower) panels are for LO (HO) of $\tz$ and the left (right) panels are for the assumed NH (IH) as the true hierarchy.  Marginalizations have been done over $\sin^2 2\ty$, $\sin^2\tz$ and $|\dmm|$.  We used a particle misidentification factor of $30\%$.}}
\label{MH-1}
\end{figure}

In general, for true NH, the hierarchy sensitivity at $\delta = 90^\circ$ is larger
than $\delta = -90^\circ$.  This is because of a larger event rate for $\delta = 90^\circ$, 
following larger probability (see Fig.~\ref{MH-1}) which arises from the CP-odd term in 
$\langle P_{e\mu}^\delta \rangle$ in Eq.~(\ref{oscCP}), resulting in smaller statistical uncertainty. The opposite is true for the case when the true hierarchy is IH. 
These can be seen immediately from our hierarchy sensitivity plots.

In Fig.~\ref{MH-1}, we have plotted the mass hierarchy sensitivity of the experimental set up described in Section~\ref{beamdetector}, after 1-year of neutrino run only (solid lines) and after $\nu$ and $\bar{\nu}$ runs of 1-year each respectively (dashed lines).  From the sensitivity curves we can see that it is possible to determine mass hierarchy for 1-year $\nu$ run only (solid lines) at more than $4\sigma$ significance, if $\tz$ lies in the LO and the true hierarchy is either NH (upper-left panel) or IH (upper-right panel). The same is true when $\tz$ is in the HO (lower panels).  The sensitivity, however, is higher when $\tz$ is in the HO and the NH is true (lower-left panel) and when $\tz$ is in the LO and the IH is true (upper-right panel) as compared to the same octant but opposite hierarchy cases.

We also see in Fig.~\ref{MH-1} that if an initial 1-year $\nu$ run is followed by a 1-year $\bar{\nu}$ run (dashed lines), it will be possible to determine the mass hierarchy at $\gtrsim 6\,\sigma$ C.L.\ for all octants of $\tz$ and true hierarchy combinations.  Note that we have marginalized over the oscillation parameters $\sin^2 2\ty$,  $\sin^2\tz$ and $|\dmm|$ to compute the sensitivity curves.

\begin{figure}[thb]
\centering
\includegraphics[width=0.48\textwidth]{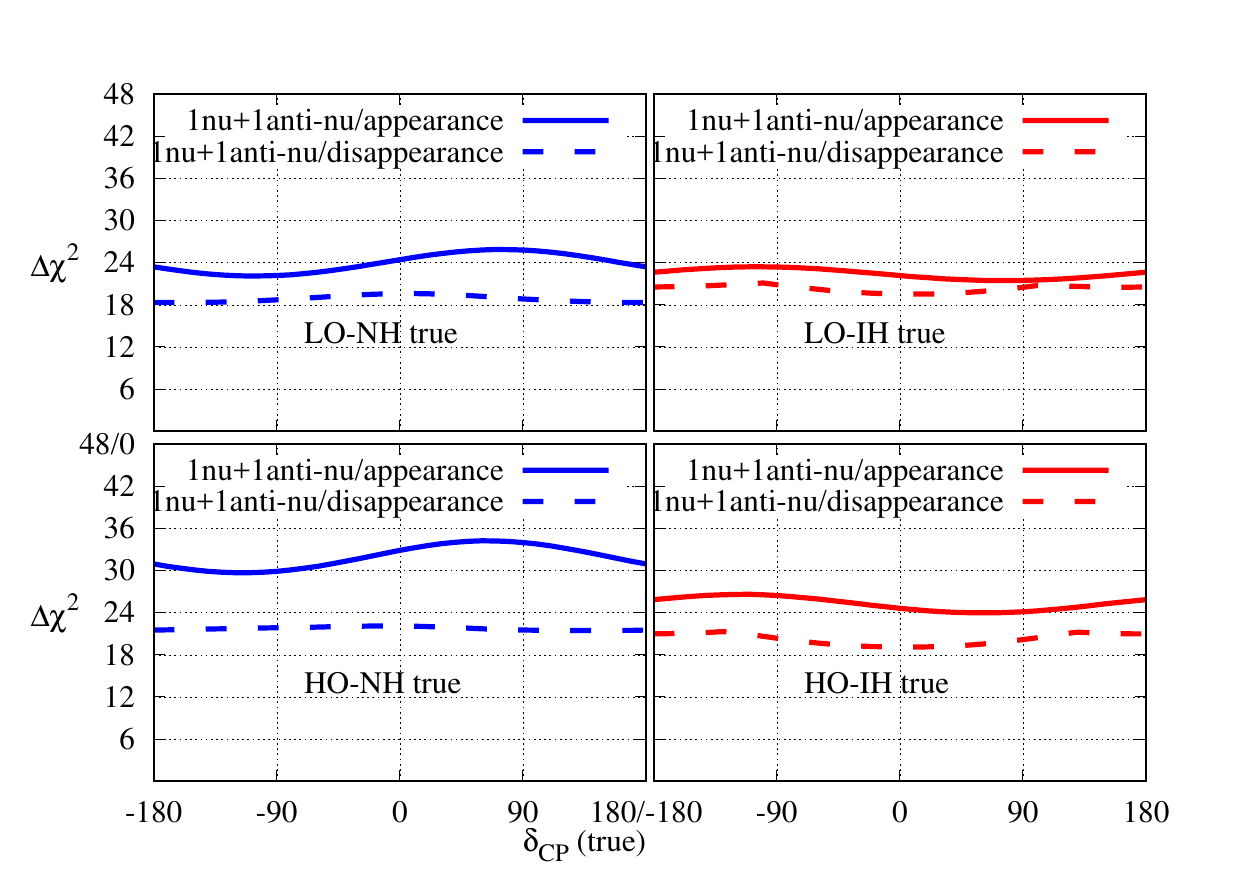}
\caption{\footnotesize{Mass hierarchy discovery potential for $\nu_\mu \rightarrow \nu_e$ oscillation channel only (appearance -- solid lines) and for $\nu_\mu \rightarrow \nu_\mu$ survival channel only (disappearance -- dashed lines), for 1-year $\nu$ 
and 1-year $\bar{\nu}$ run.  The octant of $\tz$, assumed true hierarchy, marginalizations and particle misidentification information are the same as in Fig.~\ref{MH-1}. }}
\label{MH-2}
\end{figure}

In Fig.~\ref{MH-2} we have plotted the hierarchy discovery sensitivity after 1-year $\nu$ and 
1-year $\bar{\nu}$ run, each for the $\nu_\mu \rightarrow \nu_e$ appearance/oscillation 
channel only (solid lines) and the same for the $\nu_\mu \rightarrow \nu_\mu$ disappearance/survival
channel only (dashed lines).  Note that hierarchy sensitivity is dominated by the oscillation channel  
$\nu_\mu \rightarrow \nu_e$, see Eq.~(\ref{oscHigh}).  Even if the $\nu_\mu \rightarrow \nu_\mu$ survival 
channel is ignored, it is possible to determine the hierarchy at $\gtrsim 4\,\sigma$ for all octants 
of  $\tz$ and true hierarchy.  On the other hand, the hierarchy can only be established at 
$\approx 3\,\sigma$ level in the $\nu_\mu \rightarrow \nu_\mu$ channel only (dashed lines). 
The relatively flat feature (CP-insensitive) of the dashed curves arises from the 
smallness of the CP-even term of $\langle P_{\mu\mu}\rangle$ in Eq.~(\ref{oscCP}).

\begin{figure}[thb]
\centering
\includegraphics[width=0.48\textwidth]{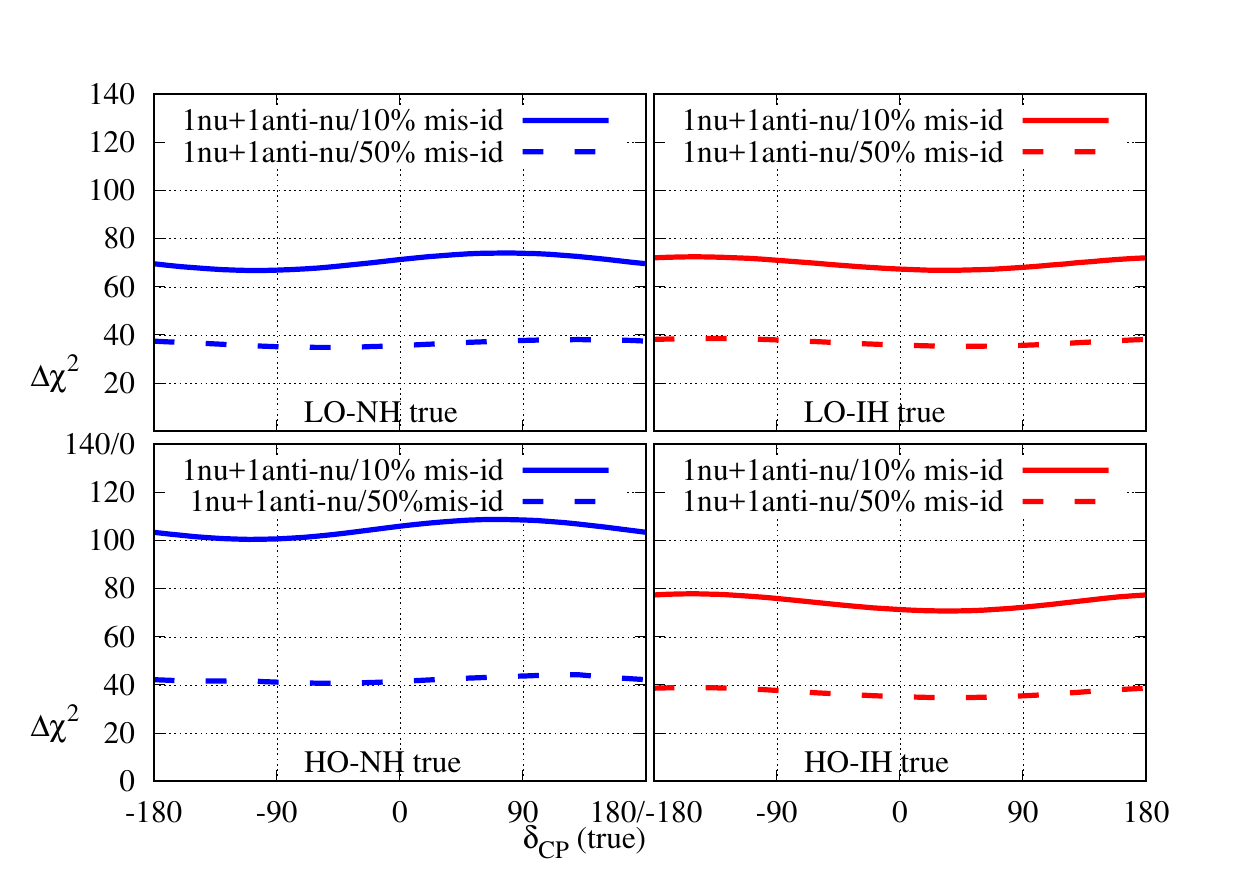}
\caption{\footnotesize{Mass hierarchy discovery potential after 1-year $\nu$ and 1-year $\bar{\nu}$ run but for particle misidentification factor $10\%$ (solid lines) and $50\%$ (dashed lines).  The octants of $\tz$, assumed true hierarchy and marginalization information are the same as in Fig.~\ref{MH-1}. }}
\label{MH-3}
\end{figure}

In Fig.~\ref{MH-3} we have explored the effects of electron-muon misidentification factors
on mass hierarchy discovery.  The solid (dashed) lines correspond to the case when
the electron-muon misidentification factor is $10\%$ ($50\%)$.  It can be seen that
when LO-NH or HO-IH is the true hierarchy-octant combination, it is possible to determine
hierarchy at $\gtrsim 8\,\sigma$ C.L.\ for $10\%$ misidentification.  
For HO-NH combination the significance 
is $\gtrsim 10\,\sigma$.  Therefore the hierarchy discovery potential for this experimental 
set up increases significantly if the electron and muon identification capability improves. 
On the other hand, it can be seen that the hierarchy discovery potential degrades drastically 
if the muon-electron identification  factor increases. When the misidentification factor is 
$50\%$, wrong hierarchy can be excluded at $\approx 6\,\sigma$ C.L.\ for all octant and true
hierarchy combinations, which is still quite good.

 \subsection{CP Phase Sensitivity}

We have studied sensitivity of ORCA, using Fermilab-LBNF beam, to measure a non-zero CP-phase value both in case of an unknown mass hierarchy and a known mass hierarchy.  As mentioned earlier, the neutrino beam peaks at an energy $\approx 2.5$ GeV and the most sensitivity to the CP-phase at the probability level is at lower energy.  Furthermore the effective mass of the ORCA detector is also rather small below $\sim 1-3$~GeV.  Both of these factors reduce sensitivity to the CP-phase.  Therefore one must consider a much longer run time.

\begin{figure}[thb]
\centering
\includegraphics[width=0.48\textwidth]{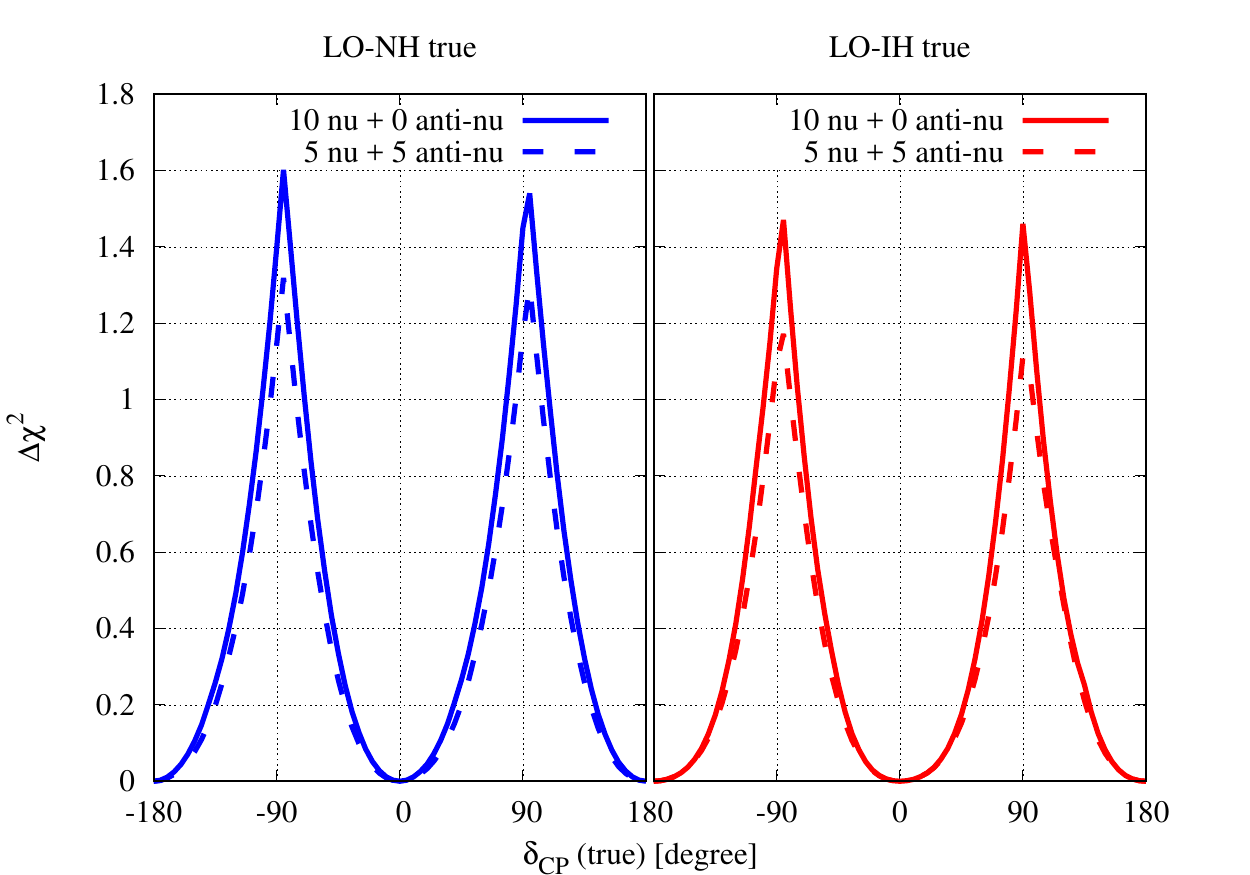}
\includegraphics[width=0.48\textwidth]{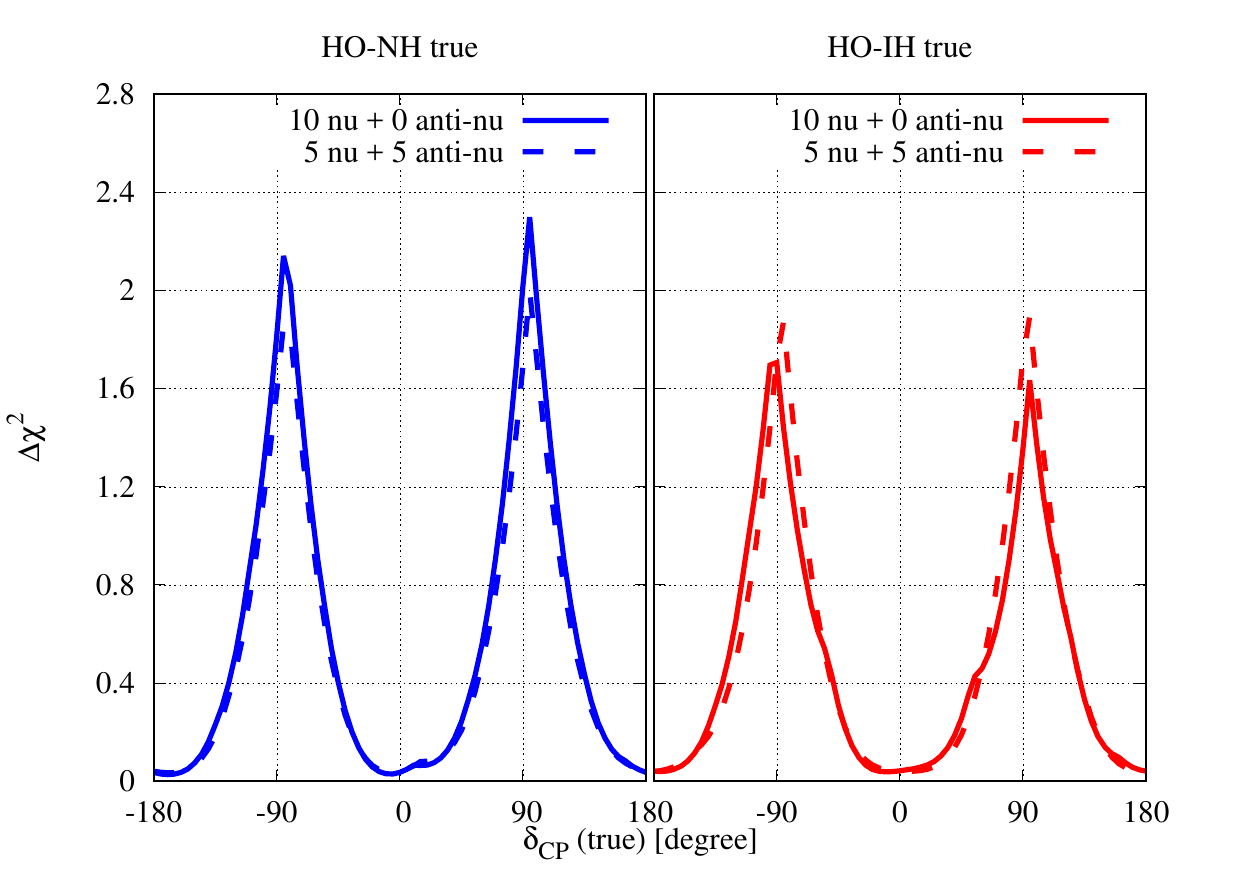}
\caption{\footnotesize{CP-violation discovery potential after 10 years of data taking. Marginalizations have been done over $\sin^2 2\ty$, 
$\sin^2\tz$, $|\dmm|$ and mass hierarchy.  We used a particle misidentification factor of $30\%$.}}
\label{CP-1}
\end{figure}

In Fig.~\ref{CP-1} we have shown the CP-phase sensitivity plots after 10-year run in the $\nu$ mode (solid lines) as well as 5-year in the $\nu$ and 5-year in the ${\bar \nu}$ mode run (dashed lines) as functions of the true $\delta$.  We have marginalized over the test values of the oscillation parameters $\sin^2 2\ty$, $\sin^2\tz$, $|\dmm|$ and mass hierarchy.  Both the muon and electron channels have been used and the electron-muon misidentification factor is $30\%$ in these plots.  Note that in case the true hierarchy is NH, the 10-year $\nu$ run gives the most sensitivity to the CP-phase (left panels - solid lines) because of a higher event rate, resulting from higher probability, and hence smaller statistical uncertainty  in the resonance channel.  The same is not true for the IH because of suppressed probability for neutrinos in this case.  Furthermore, a 5-year $\nu$ and  5-year ${\bar \nu}$ run (dashed lines) does not improve the sensitivity to the CP-phase.  This is partly because of a smaller cross-section for ${\bar \nu}$, which reduces the overall event rate, even though ${\bar P}_{\mu e}$ is enhanced for ${\bar \nu}$ in case of IH.  

We also see in Fig.~\ref{CP-1} that when NH is the true hierarchy, the sensitivity to a non-zero CP-phase is up to $\approx 1.5\,\sigma$ when $\tz$ is either in the LO or HO and for true $\delta = \pm 90^\circ$ (left panels - solid lines).  But for IH, the sensitivity is $< 1.2\,\sigma$ for all values of the true CP and for either octants of $\tz$ (right panels).

\begin{figure}[thb]
\centering
\includegraphics[width=0.48\textwidth]{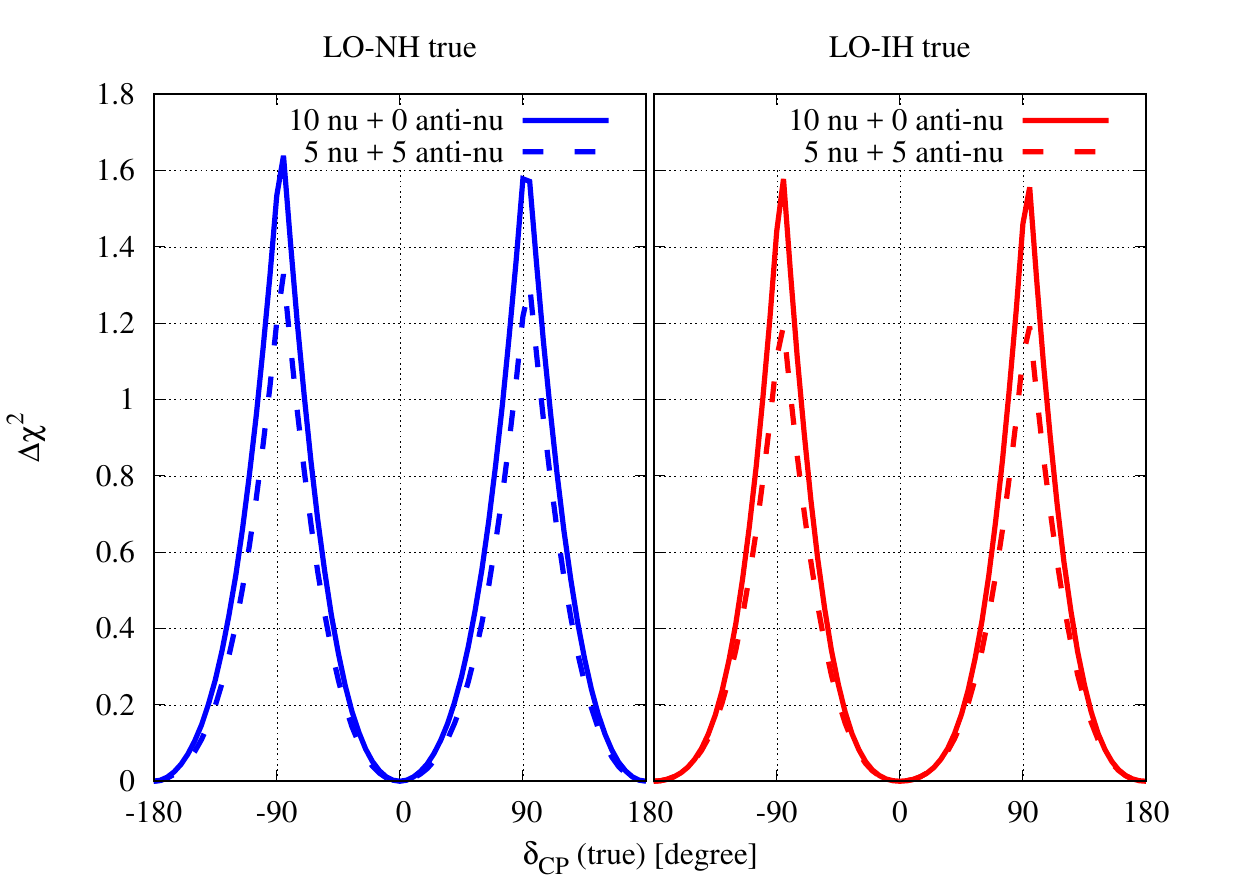}
\includegraphics[width=0.48\textwidth]{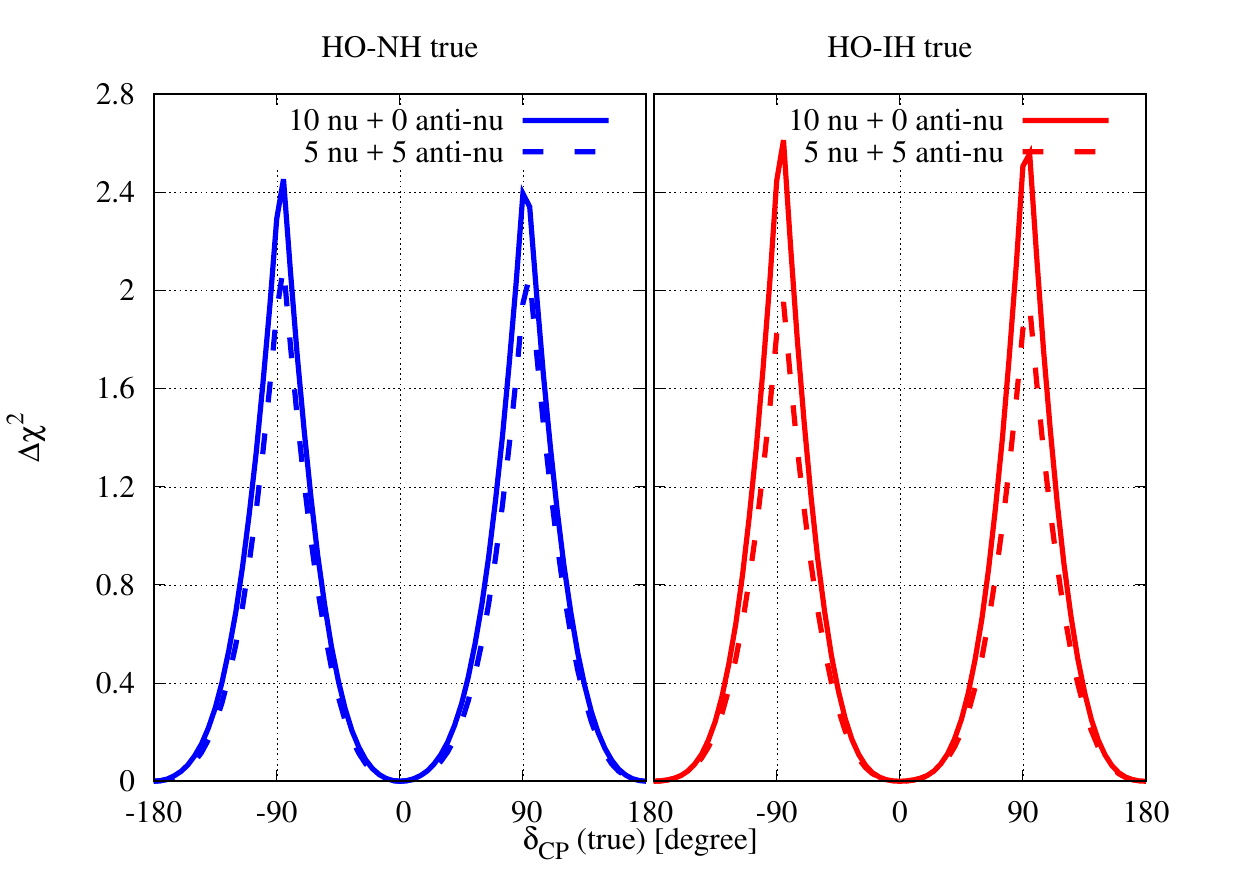}
\caption{\footnotesize{Same as Fig.~\ref{CP-1} but marginalization over $|\dmm|$ only.}}
\label{CP-2}
\end{figure}

In Fig.~\ref{CP-2} we have plotted sensitivity to the CP-phase in a case when the mass hierarchy is known (e.g., from an early run of the discussed experimental setup) and the values of $\ty$ and $\tz$ are accurately measured from other experiments.  Note that removing marginalization from $\sin^2 2\ty$, $\sin^2\tz$, and hierarchy improves significance in general but not by a large value.  This is again, due to relatively high ($\sim 1-3$~GeV) threshold energy for the proposed ORCA detector where the effect of the CP-phase on the oscillation probabilities is low.

\begin{figure}[thb]
\centering
\includegraphics[width=0.48\textwidth]{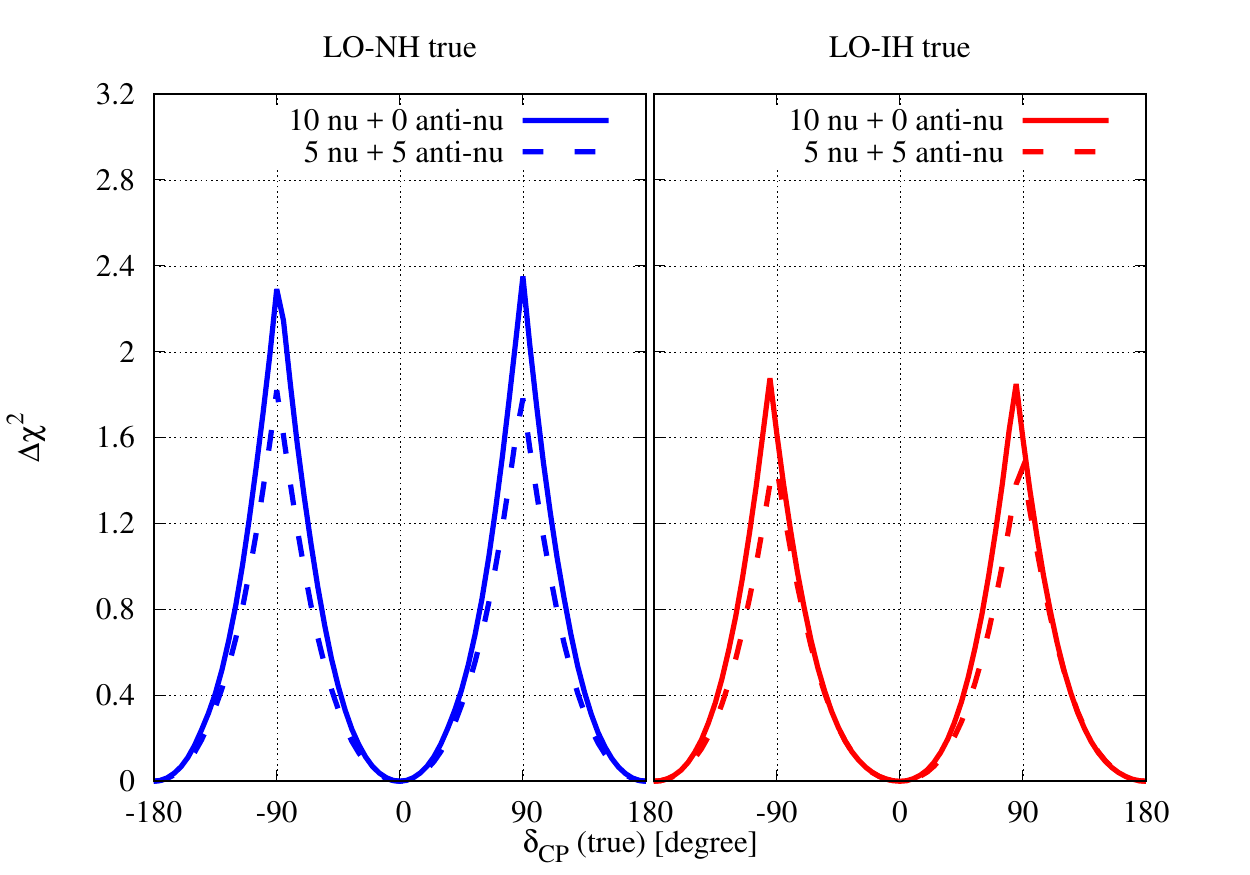}
\includegraphics[width=0.48\textwidth]{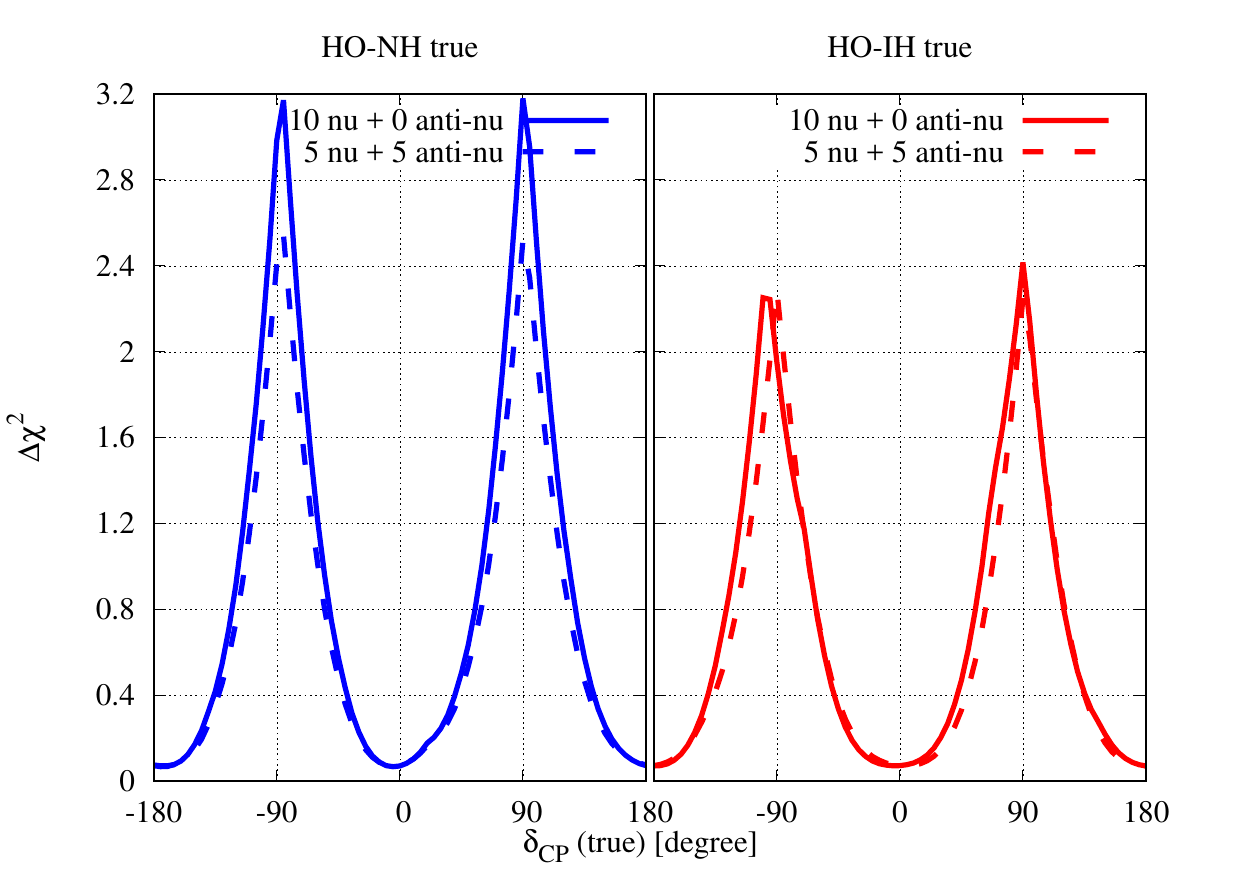}
\caption{\footnotesize{Same as Fig.~\ref{CP-1} but with $10\%$ particle misidentification.}}
\label{CP-3}
\end{figure}

We have explored the sensitivity to the CP-phase affected by the electron-muon misidentification in Fig.~\ref{CP-3}.  We used a $10\%$ misidentification factor and in general the sensitivity improves, in particular if the true hierarchy is NH, a non-zero CP-phase can be measured for $-120^\circ \lesssim \delta \lesssim -60^\circ$ and for $60^\circ \lesssim \delta \lesssim 120^\circ$ at $\gtrsim 1\,\sigma$ C.L. Marginalizations over $\sin^2 2\ty$, $\sin^2\tz$, $|\dmm|$ and mass hierarchy have been taken into account in this case.

\section{Conclusions} 
\label{conclude}

We have explored a possibility to determine the neutrino mass hierarchy and CP-phase with the proposed KM3NeT-ORCA detector in the Mediterranean sea receiving a neutrino beam from the Fermilab-LBNF, over a baseline of 6900 km.  We used publicly available detector characteristics and neutrino flux to simulate events and compute sensitivities using the GLoBES software.

Detailed efficiencies for different interaction channels in the detector have been taken into account as well as smearing of the events in energy.  We marginalized over uncertainties of the oscillation parameters and also took into account realistic particle misidentification factor in the detector.

The mass hierarchy can be determined with $\gtrsim 4\,\sigma$ significance when the true hierarchy is either normal or inverted, and for either octant of $\tz$ within 1-year of neutrino run.  A higher significance can be achieved with 1-year neutrino and 1-year antineutrino run.  For this case $\gtrsim 4\,\sigma$ significance can be reached in the electron neutrino channel only. 

The CP-phase value can only be measured in the $\pm (60^\circ - 120^\circ)$ range at $\approx 1\,\sigma$ C.L.\ for 10-year neutrino run in case the true mass hierarchy is normal.  Reducing particle misidentification in the detector can help (up to $\sim 1.8\,\sigma$ for $10\%$ misidentification) but will not increase the sensitivity significantly.

\acknowledgments

S.R. was funded by a National Research Foundation (South Africa) grant CPRR 2014 number 87823 which allowed a visit by U.R. to the University of Johannesburg where most of this work was completed.

\bibliographystyle{apsrev}

\end{document}